\documentclass[aps,prl,twocolumn,preprintnumbers,amsmath,amssymb]{revtex4-1}
\usepackage{dcolumn}
\usepackage{bm}
\usepackage{graphicx,subfigure,xcolor}
\usepackage{braket}
\usepackage{xcolor}
\usepackage[colorlinks=true,citecolor=blue,linkcolor=blue,urlcolor=blue]{hyperref}
\DeclareMathOperator{\Tr}{Tr}

\newcommand{\ad}{a^{\dag}}

\begin{document}

\title{Harvesting Multiqubit Entanglement from Ultrastrong Interactions in Circuit Quantum Electrodynamics}

\author{F. Armata$^1$}

\author{G. Calajo$^{2}$}

\author{T. Jaako$^2$}

\author{M. S.  Kim$^{1,3}$}

\author{P. Rabl$^2$}

\affiliation{$^1$QOLS and QuEST, Blackett Laboratory, Imperial College London, London SW7 2AZ, United Kingdom \\ 
$^2$Vienna Center for Quantum Science and Technology, Atominstitut, TU Wien, 1040 Vienna, Austria\\
$^3$Korea Institute of Advanced Study, Dongdaemun-gu, Seoul, 02455, South Korea}

\begin{abstract}
We analyze a multiqubit circuit QED system in the regime where the qubit-photon coupling dominates over the system's bare energy scales. Under such conditions a manifold of low-energy states with a high degree of entanglement emerges. Here we describe a time-dependent protocol for extracting these quantum correlations and converting them into well-defined multipartite entangled states of noninteracting qubits. Based on a combination of various  ultrastrong-coupling effects the protocol can be operated in a fast and robust manner, while still being consistent with experimental constraints on switching times and typical energy scales encountered in superconducting circuits. Therefore, our scheme can serve as a probe for otherwise inaccessible correlations in strongly coupled circuit QED systems. It also shows how such correlations can potentially be exploited as a resource for entanglement-based applications. 
\end{abstract}

\date{\today}

\maketitle 

Cavity QED is the study of quantum light-matter interactions with real or artificial two-level atoms  coupled to a single radiation mode. In this context one is usually interested in strong interactions between excited atomic and electromagnetic states, while  the trivial ground state, i.e.\ the vacuum state with no atomic or photonic excitations, plays no essential role. This paradigm has recently been challenged by a number of experiments \cite{Todorov2010,schwartz11,geiser12,Scalari2015,Zhang2016}, where interaction strengths comparable to the photon energy have been demonstrated.  In particular, in the field of circuit QED \cite{wallraff04,blais04}, a single superconducting two-level system can already be coupled ultrastrongly~\cite{Ciuti2005,Casanova2010,USCRemark} to a microwave resonator mode \cite{forndiaz17,yoshihara17,niemczyk10,forndiaz10,baust16, chen17,bosman17}. In this regime the physics changes drastically and even in the  ground state various nontrivial effects like spontaneous vacuum polarization \cite{nataff10a,nataff10b,Bamba2016}, light-matter decoupling \cite{deliberato14,jaako2016} and different degrees of entanglement \cite{jaako2016, levine04,hines04,ashhab10} can occur. However, compared to the vast literature on cavity QED systems in the weakly coupled regime, the opposite limit of extremely strong interactions is to a large extent still unexplored. As a consequence, ideas for how ultrastrong coupling (USC) effects can be controlled and exploited for practical applications are limited \cite{nataff11,romero12,kyaw15,wang16,wang17,stassi17}.

In this Letter we consider a prototype circuit QED system consisting of multiple flux qubits coupled to a single mode of a microwave resonator. It has recently been shown that in the USC regime this circuit exhibits a manifold of nonsuperradiant ground and low-energy  states with a high degree of multiqubit entanglement \cite{jaako2016}. This entanglement, however, is \textit{a priori} not of any particular use, since any attempt to locally manipulate or measure the individual qubits would necessarily introduce a severe perturbation to the strongly coupled system. For this reason we describe  the implementation of an entanglement-harvesting  protocol \cite{Reznik2005,Han2008,auer2011,olson2012, sabin2012, Salton2015,dai2015}, which extracts quantum correlations from USC states and converts these correlations into equivalent multipartite entangled states of decoupled qubits. The protocol combines adiabatic and nonadiabatic parameter variations and exploits the counterintuitive decoupling of qubits and photons at very strong interactions~\cite{jaako2016} to make the entanglement extraction scheme intrinsically robust and consistent with experimentally available tuning capabilities. The extracted Dicke and singlet states belong to a family of robust multipartite entangled states~\cite{Guhne2008,Bergmann2013} and form, for example,  a resource for Heisenberg-limited metrology applications~\cite{Toth2014}.  More generally, our analysis shows, how the interplay between different USC effects can contribute to the realization of non-trivial control tasks in a strongly interacting cavity QED system.

\begin{figure}[t]
\centering
\includegraphics[width=0.48\textwidth]{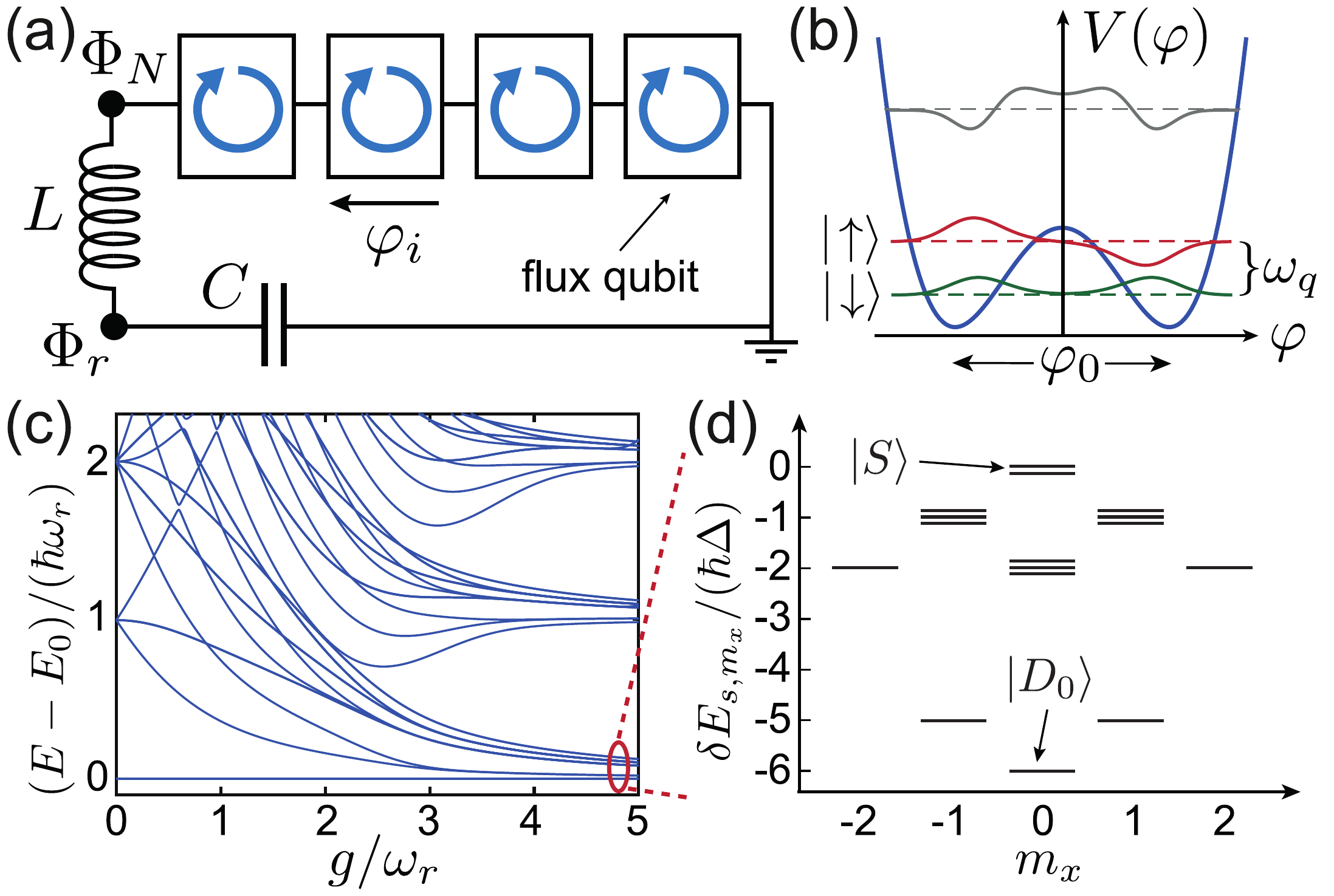}
\caption{(a) Sketch of the multiqubit circuit QED setup considered in this Letter. (b) Each flux qubit is represented by the two lowest states $\ket{\downarrow}$ and $\ket{\uparrow}$ of an effective double-well potential for the phase variable $\varphi$. Under this two-level approximation the inductive coupling $(\Phi_r-\Phi_{N})^2/(2L)$, where $\Phi_N=\Phi_0\sum_{i=1}^N \varphi_i$, gives rise to the cavity QED Hamiltonian~\eqref{Hamiltonian-quantum}. (c) Energy spectrum (with respect to the ground-state energy $E_0$) of the extended Dicke model~\eqref{DM} as a function of the coupling strength $g$ for $N=4$ and $\omega_q=\omega_r$. 
(d) Ordering of the lowest energy states in the USC regime as determined by Eq.~\eqref{DeltaE} for the case $N=4$. The multiple lines indicate the two- and threefold degeneracy of states with total angular momentum $s=0$ and $s=1$, respectively.}
\label{Fig1}
\end{figure}

\textit{Model.}---We consider a circuit QED system as shown in Fig.~\ref{Fig1}(a), where a single mode LC resonator with capacitance $C$ and inductance $L$ is coupled collectively to an even number of $N=2,4,6,\dots$ flux qubits. This circuit is described by the Hamiltonian~\cite{vool16, SuppInfo}
\begin{equation}\label{Hamiltonian}
\begin{split}
\mathcal{H}=\frac{Q_r^2}{2C}+\frac{(\Phi_r- \Phi_0 \sum_{i=1}^N  \varphi_i)^2}{2L}+\sum_{i=1}^N H_q^{(i)},
\end{split}\end{equation}
where $Q_r$ and $\Phi_r$ are charge and generalized flux operators for the resonator obeying $[\Phi_r,Q_r]=i\hbar$, and $\Phi_0=\hbar/(2e)$ is the reduced flux quantum.  
For each qubit, $H_q^{(i)}$ denotes the free Hamiltonian and $\varphi_i$ is the difference of the superconducting phase across the qubit's subcircuit. As usual we assume that the qubit dynamics can be restricted to the two lowest tunneling states $\ket{\downarrow}$ and $\ket{\uparrow}$ of a symmetric double-well potential [cf. Fig.~\ref{Fig1}(b)].  Under this approximation and writing $\Phi_r=  \sqrt{\hbar/(2C \omega_r)} (a+a^\dag)$ and $Q_r=i \sqrt{\hbar C \omega_r/2} (a^\dag-a)$, where $\omega_r=\sqrt{1/LC}$ is the resonator frequency and $a$ and $a^\dag$ are the annihilation and creation operators, we obtain 
\begin{equation}\label{Hamiltonian-quantum}
\begin{split}
\mathcal{H}=&\hbar \omega_r\ad a
+ \hbar \sum_{i=1}^N\frac{ g_i}{2}(\ad+a)\sigma^i_x\\
& +\hbar  \sum_{i=1}^N\frac{\omega_q^i}{2}\sigma_z^i + \hbar\sum_{i,j=1}^N\frac{ g_ig_j}{4\omega_r}\sigma^i_x\sigma_x^j.
\end{split}
\end{equation}
Here $\sigma_k^i$ are Pauli operators and $\omega^i_q$ are the qubit-level splittings. The second term in Eq.~\eqref{Hamiltonian-quantum} accounts for the collective qubit-resonator interaction with couplings $g_i=\Phi_0\sqrt{|\varphi_0^i|^2\omega_r/(2\hbar L)} $, where $\varphi_0^i=2\bra{\downarrow_i}\varphi_i \ket{\uparrow_i}$. The condition $g_i>\omega_r,\omega^i_q$ can be reached with an appropriate flux-qubit design~\cite{nataff10a,romero12,Bourassa2009,Peropadre2013,yoshihara17, forndiaz17}, and the $g_i(t)$ and $\omega_q^i(t)$ can be individually tuned by controlling the matrix element $\varphi_0^i$ and the height of the tunnel barrier via local magnetic fluxes \cite{romero12,sabin2012}. A specific four-junction qubit design~\cite{Peropadre2013,Qiu2016}, which combines strong interactions with a high degree of tunability, is detailed in the Supplemental Material~\cite{SuppInfo}. Finally, the last contribution in Eq.~\eqref{Hamiltonian-quantum}  represents an additional qubit-qubit interaction, which is usually neglected for cavity QED systems with weak or moderately strong couplings. However, this term is crucial in the USC regime and it is responsible for the nontrivial ground-state correlations that  are at the focus of the present Letter.

\textit{USC spectrum.}---We are primarily interested in a  symmetric configuration, i.e., $g_i=g$ and $\omega_q^i=\omega_q$. In this case the Hamiltonian~\eqref{Hamiltonian-quantum} can be expressed in terms of collective angular momentum operators $S_k=\sum_i \sigma^i_k/2$ and reduces to the extended Dicke Hamiltonian~\cite{jaako2016}
\begin{equation}\label{DM}
\mathcal{H}=\hbar \omega_r\ad a + \hbar g(\ad+a)S_x +\hbar \omega_qS_z + \hbar \frac{ g^2}{\omega_r}S_x^2.
\end{equation} 
For $g\ll \omega_r,\omega_q$ we can make a rotating wave approximation and obtain the standard Tavis-Cummings model of cavity QED with a trivial ground state $|G\rangle=|n=0\rangle\otimes \ket{\downarrow}^{\otimes N}$. If in addition $|\omega_q-\omega_r|\gg g$, all excited states are also essentially decoupled and the qubits can be individually prepared, manipulated, and measured by additional control fields. In the opposite limit, $g\gg\omega_r,\omega_q$, the coupling terms $\sim S_x$ and $\sim S_x^2$ dominate and the level structure  changes completely. This is illustrated in Fig.~\ref{Fig1}(c), which shows that for couplings $g/\omega_r\gtrsim 3$ the spectrum separates into manifolds of $2^N$ nearly degenerate states. The eigenstates in this regime are  displaced photon number states, $|\Psi_{s,m_x,n}\rangle\simeq e^{-\frac{g}{\omega_r}(a^\dag-a)S_x}|n\rangle \otimes|s,m_x\rangle$, with energies $E_{s,m_x,n}\simeq \hbar \omega_r n + \delta E^{(n)}_{s,m_x}$~\cite{jaako2016}. Here  $s$ is the total spin and $m_x=-s,\dots,s$ the spin projection quantum number; i.e., $S_x|s,m_x\rangle=m_x |s,m_x\rangle$. Within the lowest manifold, the remaining level splittings are given by
\begin{equation}\label{DeltaE}
\delta E^{(0)}_{s,m_x}=  \hbar\Delta \left[m_x^2- s(s+1)\right],\qquad \Delta= \frac{\omega_q^2 \omega_r}{2g^2},
\end{equation}
and the resulting ordering of the states is shown in Fig.~\ref{Fig1}(d) for $N=4$ qubits. Thus, for even qubit numbers $N$, the ground state in the USC regime is of the form $|\tilde G\rangle\simeq |n=0\rangle \otimes |D_0\rangle$, where $|D_{0}\rangle = |s=N/2,m_x=0\rangle$ denotes the fully symmetric Dicke state with vanishing projection along $x$. Importantly, this state exhibits a high degree of qubit-qubit entanglement, while it remains almost completely decoupled from the cavity field~\cite{jaako2016}. Our goal is now to identify a suitable protocol for converting this state into an equivalent state of the decoupled system, where it becomes available as an entanglement resource for further use.

\begin{figure}
\centering
\includegraphics[width=0.48\textwidth]{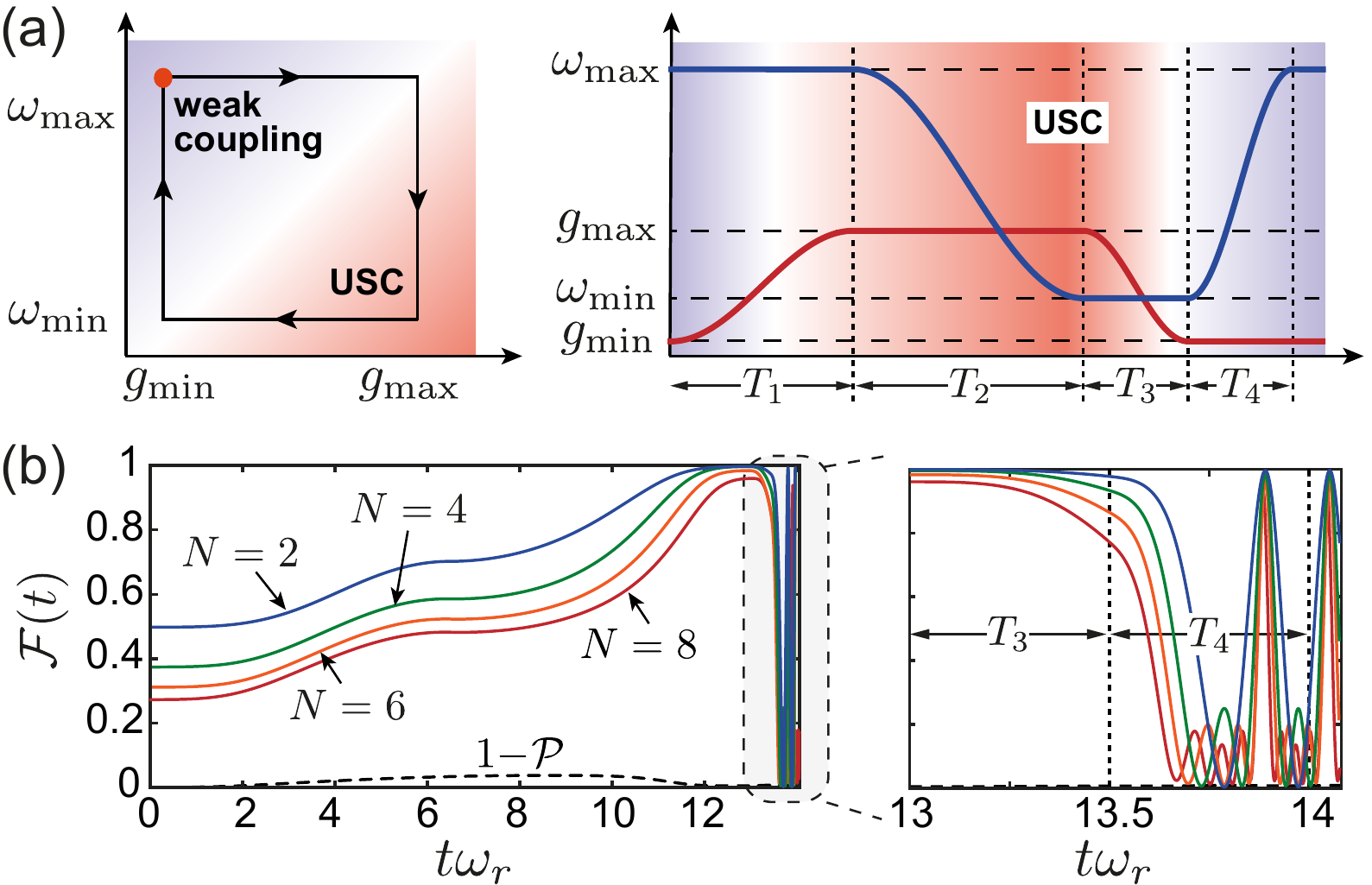}
\caption{(a) General pulse sequence for the qubit parameters $\omega_q(t)$ and $g(t)$ considered for the implementation of the entanglement harvesting protocol. (b) The fidelity $\mathcal{F}(t)$ is plotted as a function of time and for different qubit numbers. The dashed line indicates 
the quantity $1-\mathcal{P}(t)$, where $\mathcal{P}(t)={\rm Tr}\{\rho_q^2(t)\}$ is the purity of the reduced qubit state $\rho_q(t)={\rm Tr}_r\{\rho(t)\}$ for the case $N=4$. It shows that after an intermediate stage of finite qubit-resonator entanglement, the purity of the qubit state is almost fully restored when the system enters deep into the USC regime.
For all values of $N$ the same parameters $\omega_{\rm max}/\omega_r=20$, $\omega_{\rm min}/\omega_r=0.5$, $g_{\rm max}/\omega_r=4.5$, $g_{\rm min}/\omega_r=0.1$ and times intervals $T_1=T_2=6.5\omega_r^{-1}$ and $T_3=T_4=0.5\omega_r^{-1}$ have been assumed.} 
\label{Fig2}
\end{figure}

\textit{Entanglement harvesting.}---Figure~\ref{Fig2}(a) shows a general pulse sequence for implementing the  entanglement-harvesting protocol through variations of  $\omega_q(t)$ and $g(t)$. For this protocol the system is initialized in the ground state $|G\rangle$ of the weakly coupled system, where the qubits are far detuned from the cavity, $\omega_q=\omega_{\rm max}\gg\omega_r$, and the coupling is set to a minimal value, $g=g_{\rm min}<\omega_{r}$. In the first two steps, $T_1$ and $T_2$, the system is adiabatically tuned into the USC regime with a maximal coupling $g_{\rm max}>\omega_r$ and a low value of the qubit frequency $\omega_{\rm min}\lesssim \omega_r$. This process prepares the system in the USC ground state $|\tilde G\rangle$. In the successive steps, $T_3$ and $T_4$, the qubits and the resonator mode are separated again, but now in the reverse order and using nonadiabatic parameter variations.  Ideally, during this part of the protocol the system simply remains in state $|\tilde G\rangle$ and becomes the desired excited state of the weakly coupled system at the final time $T_{\rm f}=\sum_{n=1}^4 T_n$. This general sequence achieves two main goals. First, the adiabatic preparation stage can be implemented very rapidly, since it must only be slow compared to the fast time scales set by $\omega_{\rm max}^{-1}$ and $g_{\rm max}^{-1}$. At the same time the nonadiabatic decoupling processes only need to be fast compared to the slow time scales $\omega^{-1}_r$, $\omega^{-1}_{\rm min}$, and $g_{\rm min}^{-1}$. This second condition is most crucial for a time-dependent control of USC systems, since it makes the required switching times experimentally accessible and consistent with the two-level approximation assumed in our theoretical model.

In Fig.~\ref{Fig2}(b) we plot the fidelity $\mathcal{F}(t)=\Tr\{\rho(t) |D_0\rangle\langle D_0|\}$, where $\rho(t)$ is the density operator of the full system, for a specific set of pulse parameters listed in the figure caption. We see that the entanglement extraction fidelity (EEF) $\mathcal{F}_{\rm E} ={\rm max}\{\mathcal{F}(t)| t \geq T_{\rm f}\}$, i.e., the maximal fidelity after the decoupling step, reaches near perfect values of $ \mathcal{F}_{\rm E} \simeq 0.95-0.99$ for different numbers of qubits, without any further fine-tuning of the control pulses.  Note that the fidelity oscillations at the end of the sequence are simply due to the fact that $|D_0\rangle$ is not an eigenstate of the bare qubit Hamiltonian, $H_q=\omega_q S_z$. However, this evolution does not affect the purity or the degree of entanglement of the final qubit state and can be undone by local qubit rotations.

\begin{figure}
\centering
\includegraphics[width=0.48\textwidth]{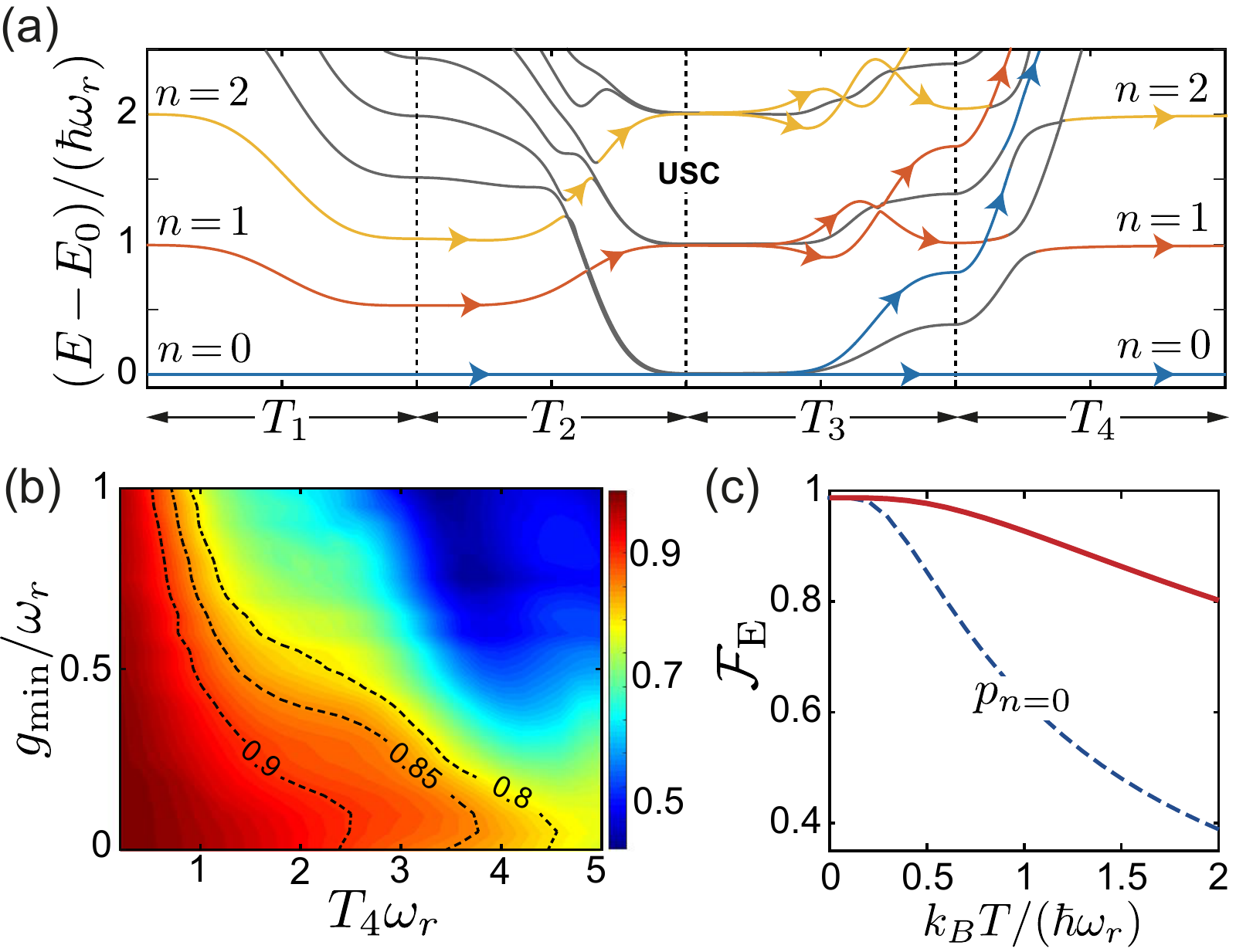}
\caption{(a) Evolution of the lowest eigenvalues during different stages of the protocol for the case $N=2$. Here $g_{\rm min}/\omega_r=0.2$, $\omega_{\rm min}/\omega_r=0.4$, and in the final step of the protocol $\omega_{\rm max}/\omega_r=5$. For clarity only the $s=1$ states are shown and all time intervals have been stretched to equal lengths. For different initial photon number states $|n\rangle$, the colored segments and arrows indicate the ideal evolution of the systems, which maximizes the probability to end up in the qubit state $|D_0\rangle=(\ket{\uparrow\uparrow}-\ket{\downarrow\downarrow})/\sqrt{2}$. Nonadiabatic crossings occur during the fast decoupling steps ($T_3$ and $T_4$), but also for small avoided crossings in the excited state manifolds during the preparation step ($T_2$).   (b)  Plot of the EEF for varying $T_4(=T_3)$ and $g_{\rm min}$ and for $N=4$. (c)  EEF (solid line) for a resonator mode, which is initially in a thermal state at temperature $T$, for $N=4$. The dashed line indicates the corresponding population of the ground state manifold. All the other pulse parameters in panels (a), (b) and (c) are the same as in Fig.~\ref{Fig2}(b).} 
\label{Fig3}
\end{figure}

\emph{Experimental considerations.}---For a possible experimental implementation of the protocol we consider qubits with a frequency of $\omega_{\rm max}/(2\pi)\approx10$ GHz coupled to a lumped-element  resonator of frequency $\omega_r/(2\pi)=500$ MHz. The required maximal coupling strength of $g_{\rm max}\simeq 4.5\omega_r \approx 2\pi \times 2.25$  GHz is then consistent with experimentally demonstrated values~\cite{forndiaz17, yoshihara17}. For these parameters, the nonadiabatic switching times assumed in Fig.~\ref{Fig2}(b) correspond to $T_{3,4}\simeq 0.16$ ns. These switching times are within reach of state-of-the-art waveform generators and a sinusoidal modulation of flux qubits on such time scales has already been demonstrated~\cite{Wilson2011}. At the same time the duration of the whole protocol, $T_{\rm f}=15/\omega_r\approx 5$ ns, is still much faster than typical flux qubit coherence times of  1-100 $\mu$s~\cite{Yan2016} or the lifetime of a photon, $T_{\rm ph}=Q/\omega_r$, in a microwave resonator of quality factor $Q=10^4-10^6$.  Therefore, although many experimental techniques for implementing and operating circuit QED systems in the USC regime are still under development, these estimates clearly demonstrate the feasibility of realizing high-fidelity control operations in such devices.

In practice additional limitations might arise from the lack of complete tunability of $g(t)$ and $\omega_q(t)$. This is illustrated in Fig.~\ref{Fig3}(a), which shows the evolution of the lowest eigenenergies during different stages of the protocol for the case $N=2$ and a nonvanishing value of $g_{\rm min}$. In this case the appearance of several avoided crossings during the final ramp-up step prevents a fully nonadiabatic decoupling. In Fig.~\ref{Fig3}(b) we plot the resulting EEF for varying $g_{\rm min}$ and $T_4$. This plot demonstrates the expected trade-off between the residual coupling and the minimal switching time, but also that the protocol is rather robust and fidelities of $\mathcal{F}_{\rm E}\sim 0.9$ are still possible for minimal  couplings of a few hundred MHz or switching times approaching $\sim 1$ ns. 
Similar conclusions are obtained when a partial dependence between the pulses for $g(t)$ and $\omega_q(t)$ or nonuniform couplings $g_i(t)$ and frequencies $\omega_q^i(t)$ due to fabrication uncertainties are taken into account. Numerical simulations of the protocol under such realistic experimental conditions~\cite{SuppInfo} demonstrate that no precise fine-tuning of the system parameters is required.

\emph{Extracting entanglement from a thermal state.}---The above-considered  protocol relies on a rather low resonator frequency $\omega_r$ in order to enhance both $g/\omega_r$ as well as  the nonadiabatic switching times. This implies that even at temperatures of $T=20$ mK the equilibrium populations of higher resonator states with $n\geq 1$ cannot be neglected.  In Fig.~\ref{Fig3}(c) we plot the EEF as a function of the  temperature $T$, assuming an initial resonator state $\rho_{\rm th}=\sum_n p_n\ket{n}\bra{n}$, where $p_n=\bar{n}^n/(1+\bar{n})^{n+1}$ is the thermal distribution for a mean excitation number $\bar{n}=1/(e^{\hbar\omega_r/k_BT}-1)$. We see that the EEF is significantly higher than one would naively expect from the initial population in the ground state $|G\rangle$. The origin of this surprising effect can  be understood from the eigenvalue plot in Fig.~\ref{Fig3}(a). For example, the weak-coupling eigenstate $|n=1\rangle\otimes \ket{\downarrow}^{\otimes N}$ is efficiently mapped on the corresponding USC state $|n=1\rangle\otimes|s=N/2,m_x=0\rangle$, passing only through a weak, higher-order avoided crossing. Therefore, the intermediate---and as a result also the final---qubit state is one with the resonator being in state $\ket{1}$. Although for higher photon numbers the avoided crossings become more relevant, the protocol still approximately implements the mapping $|n\rangle\otimes \ket{\downarrow}^{\otimes N}\rightarrow |n\rangle\otimes|s=N/2,m_x=0\rangle$, independent of the resonator state $\ket{n}$.  This feature makes it rather insensitive to thermal occupations and avoids additional active cooling methods for initializing the system in state $\ket{G}$.

\begin{figure}
\centering
\includegraphics[width=0.48\textwidth]{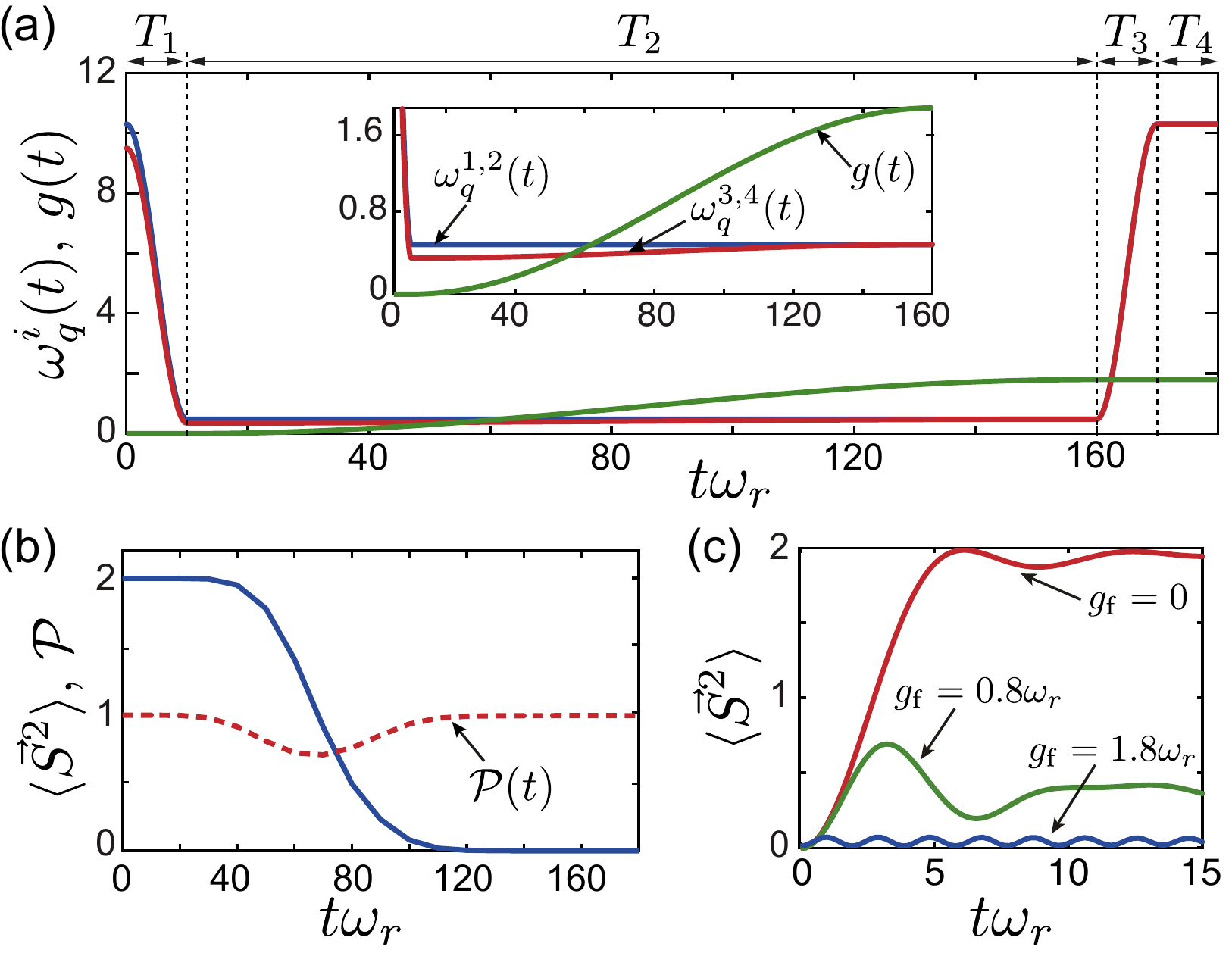}
\caption{(a) Pulse sequence for harvesting the 4-qubit entangled state $|S\rangle$ with total angular momentum $s=0$. As shown in the inset, during the first part of the protocol a finite difference between the qubit frequencies $\omega_q^{1,2} $ and $\omega_q^{3,4} $ is used to break the symmetry and couple different angular momentum states. (b) The expectation value of the total spin, $\langle \vec S^2(t) \rangle$, (solid line) and the purity of the reduced qubit state, $\mathcal{P}(t)$, (dashed line) are plotted for the pulse sequence shown in (a) and for an initial state $|\Psi_0\rangle=\ket{0}\otimes\ket{\uparrow \uparrow \downarrow \downarrow} $. (c) Evolution of the extracted state $|0\rangle\otimes|S\rangle$ (characterized by the expectation value of the total spin) after the protocol for different final values of the couplings $g_{\rm f}$. For this plot an average over random distributions of the qubit frequencies, $\omega_q^i=\omega_q(1+\epsilon_i)$, has been assumed, where $\omega_q/\omega_r=10$ and the $\epsilon_i$ are  chosen randomly from the interval $[-0.05,0.05]$. For very strong couplings, the residual oscillations indicate that all transitions induced by the nonuniform $\epsilon_i$ from state $|S\rangle$ to other states are highly detuned.} 
\label{Fig4}
\end{figure}

\emph{Entanglement protection.}---Figure~\ref{Fig1}(d) shows that apart from the ground state $|\tilde G\rangle$ there are many other highly entangled states within the lowest USC manifold. Of particular interest is the energetically highest state  $\ket{\tilde E}=\ket{n=0}\otimes|S\rangle$, where $|S\rangle$ is a singlet state with total angular momentum $s=0$ and $S_z|S\rangle=S_x|S\rangle=0$. Therefore, once prepared, this state is an exact dark state of Hamiltonian~\eqref{DM} and remains decoupled from the cavity field in all parameter regimes. Although this state is not connected to any of the bare qubit states in a simple adiabatic way, it can still be harvested by an adopted protocol, as described in Fig.~\ref{Fig4}(a) for the case $N=4$. For this protocol the system is initially prepared in the excited state $|\Psi_0\rangle=|0\rangle\otimes\ket{\uparrow \uparrow\downarrow\downarrow}$ and in a first step the qubit states are lowered below the resonator frequency in order to avoid further level crossings with higher-photon-number states. The increase of the coupling combined with a frequency offset to break the angular momentum conservation then evolves the system into a state with $s=0$ already for moderate couplings of $g/\omega_r\approx 1.8$. Note that for $N\geq 4$ there are  multiple degenerate USC states with $s=0$~\cite{Dicke1954,Arecchi1972} [cf. Fig.~\ref{Fig1}(d)], out of which the protocol selects a specific superposition~\cite{SuppInfo}.

Although the harvesting protocol for state $\ket{S}$ loses some of the robustness of the ground-state protocol, it adds an important feature. By retaining a finite coupling $g_{\rm f}=g(t=T_{\rm f})\sim \omega_r$ at the end of the protocol, the extracted dark state $|S\rangle$ is energetically separated from all other states with $s\neq0$ and it is thereby protected against small frequency fluctuations. This effect is illustrated in Fig.~\ref{Fig4}(c), which shows the evolution of the extracted state $|S\rangle$ in the presence of small random shifts of the individual qubit frequencies. For $g_{\rm f}=0$ this leads to dephasing of the qubits and a rapid transition out of the $s=0$ subspace. This dephasing can be substantially suppressed by keeping the coupling at a finite value. Thus, this example shows that USC effects can be used not only to generate complex multiqubit entangled states, but also to protect them. 

\textit{Conclusion.}---We have presented a  protocol for extracting well-defined multiqubit-entangled states from the ground-state manifold of an ultrastrongly coupled circuit QED system. The detailed analysis of this protocol illustrates, how various--so far unexplored--USC effects can contribute to a robust generation and protection of complex multiqubit states. These principles can serve as a guideline for many other preparation, storage and control operations in upcoming USC circuit QED experiments with two or more qubits. 

\textit{Acknowledgments.}---This work was supported by the Austrian Science Fund (FWF) through the SFB FoQuS, Grant No. F40, the DK CoQuS, Grant No. W 1210, and the START Grant No. Y 591-N16, and the People Programme (Marie Curie Actions) of the European Union's Seventh Framework Programme (FP7/2007-2013) under REA Grant No. 317232. F. A. wishes to express his gratitude to the Quantum Optics Theory Group of Atominstitut (TU Wien) for the warm hospitality received during his numerous visits. M. S. K. acknowledges support from the UK EPSRC Grant No. EP/K034480/1 and the Royal Society.

\widetext

\clearpage
\setcounter{equation}{0}
\setcounter{figure}{0}
\setcounter{table}{0}
\setcounter{page}{6}
\makeatletter
\renewcommand{\theequation}{S\arabic{equation}}
\renewcommand{\thefigure}{S\arabic{figure}}
\renewcommand{\bibnumfmt}[1]{[S#1]}
\renewcommand{\citenumfont}[1]{S#1}

\begin{center}
    \textbf{\large Supplemental material}
\end{center}

\section{USC eigenstates}
In the USC regime, the eigenstates of the extended Dicke Hamiltonian are labeled by the total spin $s=0,1,\dots,N/2$ and the spin projection quantum number $m_x=-s,\dots,s$, i.e., $\vec S^2|s,m_x\rangle=s(s+1)|s,m_x\rangle$ and $S_x|s,m_x\rangle=m_x |s,m_x\rangle$. For $N>2$ the states $|s\neq N/2,m_x\neq\pm N/2\rangle$ appear as multiplets~\cite{Arecchi1972s}, due to the permutation symmetry of the Hamiltonian.  In the following we provide an overview of the relevant spin states for the case of $N=2$ and $N=4$ qubits. These states are most conveniently expressed in the rotated basis 
\begin{equation}
\ket{\downarrow}_x=\frac{1}{\sqrt{2}}\left(\ket{\downarrow}-\ket{\uparrow}\right),\qquad \ket{\uparrow}_x=\frac{1}{\sqrt{2}}\left(\ket{\downarrow}+\ket{\uparrow}\right).
\end{equation}

\subsection{2 qubits}
For two qubits we have the usual three triplet states
\begin{equation}\label{s=1}
\begin{split}
&\ket{s=1,m_x=1}=\ket{\uparrow\uparrow}_x,\\
&\ket{s=1,m_x=0}=\frac{1}{\sqrt{2}}(\ket{\uparrow\downarrow}_x+\ket{\downarrow\uparrow}_x),\\
&\ket{s=1,m_x=-1}=\ket{\downarrow\downarrow}_x,
\end{split}
\end{equation}
and the singlet 
\begin{equation}\label{s=0}
\begin{split}
\ket{s=0,m_x=0}=\frac{1}{\sqrt{2}}(\ket{\uparrow\downarrow}_x-\ket{\downarrow\uparrow}_x).
\end{split}
\end{equation}
When expressed in terms of the original qubit basis the two $m_x=0$ states of interest read
\begin{equation}
\ket{s=1,m_x=0}=\frac{1}{\sqrt{2}}(\ket{\downarrow\downarrow}-\ket{\uparrow\uparrow}),\qquad \ket{s=0,m_x=0}=\frac{1}{\sqrt{2}}(\ket{\uparrow\downarrow}-\ket{\downarrow\uparrow}).
\end{equation}

\subsection{4 qubits}
For the case of $N=4$ qubits we obtain  a quintuplet for $s=2$, three triplets for $s=1$ and a two states for $s=0$. The maximally symmetric states with $s=2$ are the usual Dicke states in the $x$-basis, i.e.,  
\begin{equation}\label{s=2}
\begin{split}
&\ket{s=2,m_x=2}=\ket{\uparrow\uparrow\uparrow\uparrow}_x,\\
&\ket{s=2,m_x=1}=\frac{1}{2}(\ket{\downarrow\uparrow\uparrow\uparrow}_x+\ket{\uparrow\downarrow\uparrow\uparrow}_x+\ket{\uparrow\uparrow\downarrow\uparrow}_x+\ket{\uparrow\uparrow\uparrow\downarrow}_x),\\
&\ket{s=2,m_x=0}=\frac{1}{\sqrt{6}}(\ket{\downarrow\downarrow\uparrow\uparrow}_x+\ket{\uparrow\uparrow\downarrow\downarrow}_x+\ket{\downarrow\uparrow\uparrow\downarrow}_x+\ket{\uparrow\downarrow\downarrow\uparrow}_x+\ket{\downarrow\uparrow\downarrow\uparrow}_x+\ket{\uparrow\downarrow\uparrow\downarrow}_x),\\
&\ket{s=2,m_x=-1}=\frac{1}{2}(\ket{\downarrow\uparrow\downarrow\downarrow}_x+\ket{\downarrow\downarrow\downarrow\uparrow}_x+\ket{\downarrow\downarrow\uparrow\downarrow}_x+\ket{\uparrow\downarrow\downarrow\downarrow}_x),\\
&\ket{s=2,m_x=-2}=\ket{\downarrow\downarrow\downarrow\downarrow}_x.\\
\end{split}
\end{equation}
For the entanglement harvesting protocol, we are interested in the state $|s=2,m_x=0\rangle$, which in the original qubit basis is given by  
\begin{equation}
\ket{s=2,m_x=0}=\frac{3}{\sqrt{24}}(\ket{\uparrow\uparrow\uparrow\uparrow}+\ket{\downarrow\downarrow\downarrow\downarrow})-\frac{1}{\sqrt{24}}(\ket{\uparrow\uparrow\downarrow\downarrow}+\ket{\uparrow\downarrow\uparrow\downarrow}+\ket{\uparrow\downarrow\downarrow\uparrow}+\ket{\downarrow\uparrow\uparrow\downarrow}+\ket{\downarrow\uparrow\downarrow\uparrow}+\ket{\downarrow\downarrow\uparrow\uparrow}).
\end{equation}
Each of the $s=1$ states is 3-fold degenerate and the corresponding states are 
\begin{equation}\label{s=1}
\begin{split}
\ket{s=1,m_x=1}&=\begin{cases}
    \frac{1}{2}(\ket{\uparrow\uparrow\uparrow\downarrow}_x+\ket{\uparrow\uparrow\downarrow\uparrow}_x-\ket{\uparrow\downarrow\uparrow\uparrow}_x-\ket{\downarrow\uparrow\uparrow\uparrow}_x)\\
     \frac{1}{2}(\ket{\uparrow\uparrow\uparrow\downarrow}_x-\ket{\uparrow\uparrow\downarrow\uparrow}_x+\ket{\uparrow\downarrow\uparrow\uparrow}_x-\ket{\downarrow\uparrow\uparrow\uparrow}_x)\\
     \frac{1}{2}(\ket{\uparrow\uparrow\uparrow\downarrow}_x-\ket{\uparrow\uparrow\downarrow\uparrow}_x-\ket{\uparrow\downarrow\uparrow\uparrow}_x+\ket{\downarrow\uparrow\uparrow\uparrow}_x)\\
  \end{cases}\\
\ket{s=1,m_x=0}&=\begin{cases}
    \frac{1}{\sqrt{2}}(\ket{\uparrow\uparrow\downarrow\downarrow}_x-\ket{\downarrow\downarrow\uparrow\uparrow}_x)\\
  \frac{1}{\sqrt{2}}(\ket{\uparrow\downarrow\uparrow\downarrow}_x-\ket{\downarrow\uparrow\downarrow\uparrow}_x)\\
    \frac{1}{\sqrt{2}}(\ket{\uparrow\downarrow\downarrow\uparrow}_x-\ket{\downarrow\uparrow\uparrow\downarrow}_x)\\
  \end{cases}\\
\ket{s=1,m_x=-1}&=\begin{cases}
\frac{1}{2}(\ket{\uparrow\downarrow\downarrow\downarrow}_x+\ket{\downarrow\uparrow\downarrow\downarrow}_x-\ket{\downarrow\downarrow\uparrow\downarrow}_x-\ket{\downarrow\downarrow\downarrow\uparrow}_x)\\
    \frac{1}{2}(\ket{\uparrow\downarrow\downarrow\downarrow}_x-\ket{\downarrow\uparrow\downarrow\downarrow}_x+\ket{\downarrow\downarrow\uparrow\downarrow}_x-\ket{\downarrow\downarrow\downarrow\uparrow}_x)\\
    \frac{1}{2}(\ket{\uparrow\downarrow\downarrow\downarrow}_x-\ket{\downarrow\uparrow\downarrow\downarrow}_x-\ket{\downarrow\downarrow\uparrow\downarrow}_x+\ket{\downarrow\downarrow\downarrow\uparrow}_x)\\
  \end{cases}
\end{split}
\end{equation}

Finally, for $N=4$ we obtain two singlet states with $s=0$, which are given by
\begin{equation}\label{s=0}
\begin{split}
\ket{s=0,m_x=0}&=\begin{cases}
|S\rangle=  \frac{1}{\sqrt{3}}(\ket{\uparrow\uparrow\downarrow\downarrow}_x+\ket{\downarrow\downarrow\uparrow\uparrow}_x)-\frac{1}{\sqrt{12}}(\ket{\uparrow\downarrow\uparrow\downarrow}_x+\ket{\uparrow\downarrow\downarrow\uparrow}_x+\ket{\downarrow\uparrow\uparrow\downarrow}+\ket{\downarrow\uparrow\downarrow\uparrow}_x),\\
|S^\prime\rangle= \frac{1}{2}(\ket{\uparrow\downarrow\uparrow\downarrow}_x-\ket{\uparrow\downarrow\downarrow\uparrow}_x-\ket{\downarrow\uparrow\uparrow\downarrow}_x+\ket{\downarrow\uparrow\downarrow\uparrow}_x) .
  \end{cases}
\end{split}
\end{equation}
Since the $s=0$ subspace is invariant under rotations, the states have the same form as in the original qubit basis,
\begin{equation}\label{s=0,z}
\begin{split}
\ket{s=0,m_x=0}&=\begin{cases}
|S\rangle= \frac{1}{\sqrt{3}}(\ket{\uparrow\uparrow\downarrow\downarrow}+\ket{\downarrow\downarrow\uparrow\uparrow})-\frac{1}{\sqrt{12}}(\ket{\uparrow\downarrow\uparrow\downarrow}+\ket{\uparrow\downarrow\downarrow\uparrow}+\ket{\downarrow\uparrow\uparrow\downarrow}+\ket{\downarrow\uparrow\downarrow\uparrow}),\\
|S^\prime\rangle=\frac{1}{2}(\ket{\uparrow\downarrow\uparrow\downarrow}-\ket{\uparrow\downarrow\downarrow\uparrow}-\ket{\downarrow\uparrow\uparrow\downarrow}+\ket{\downarrow\uparrow\downarrow\uparrow}).
  \end{cases}
\end{split}
\end{equation}
Note that in Eqs.~\eqref{s=0} and \eqref{s=0,z} the specific choice of basis states has been used to match the state $|S\rangle$ generated in the protocol described in Fig. 4 in the main text and in the following section.

\begin{figure}
\includegraphics[width=0.65\textwidth]{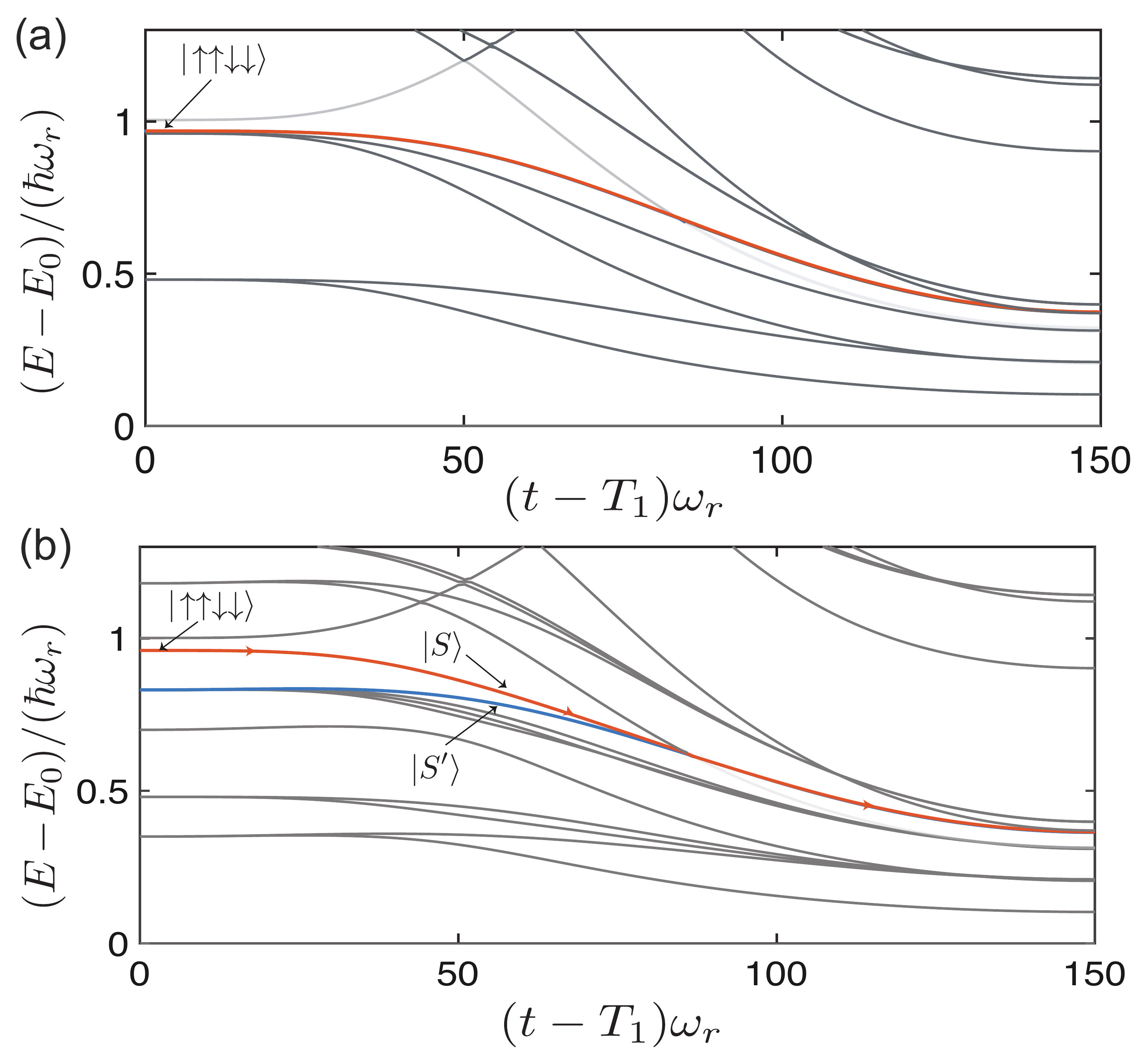}
\caption{Evolution of the  eigenvalues during the second step of the protocol for extracting the state $|S\rangle$ [see Fig 4 in the main text]. The colored lines indicate the states, which are adiabatically converted into the two $s=0$ states, $|S\rangle$ and $|S^\prime\rangle$. In plot (a) all qubit frequencies are the same, $\omega_q^i=0.48\omega_r$, while in plot (b) values $\omega_q^{1,2}=0.48\omega_r$ and $\omega_q^{3,4}=0.35\omega_r$ have been assumed. In both plots the couplings $g_i=g$ are raised symmetrically during the period $T_2$, i.e., the plotted time span, from the value $g_{\rm min}=0$ to the value $g_{\rm max}=1.8\omega_r$. Note that for these modest coupling values, the state $|S\rangle$ is not yet the energetically highest state in the $n=0$ manifold [compare Fig. 1(c) in the main text].}
\label{fig_sup_singlet}
\end{figure}

\section{Protocol for the generation of the singlet qubit states with $s=0$}

Compared to the state $|D_0\rangle$ the singlet states $|s=0,m_x=0\rangle$, defined by $\vec S^2|s=0, m_x=0\rangle=S_z |s=0, m_x=0\rangle=S_x|s=0, m_x=0\rangle=0$, are not directly adiabatically connected to any of the bare states. Nevertheless these states can still be prepared by using an adapted protocol (Fig.~4 of the main text) that we are going to describe here in more detail. 

Similar to the ground-state protocol, we start from the decoupled regime where $g\simeq 0$ and $\omega_q^i\gg \omega_r$, but we initialize the system in the excited qubit state $|\Psi_0\rangle=|0\rangle\otimes\ket{\uparrow \uparrow \downarrow \downarrow}$ ($|\Psi_0\rangle=|0\rangle\otimes\ket{\uparrow \downarrow}$ for $N=2$ ), where half of the qubits are in the excited state and half in the ground state. Note that for $N=4$ and a fully symmetric system, the $s=0$ manifold is two-fold degenerate and spanned, e.g., by the basis states given in Eq.~\eqref{s=0,z}. To prepare a well-defined state, we break the symmetry by creating an offset between the qubit frequencies, for example, by setting $\omega_q^{1,2}\ne\omega_q^{3,4}$. Once the state $|\Psi_0\rangle$ is prepared, all the qubit frequencies are lowered below the resonator frequency such that $\omega_q^i<\omega_r/2$. This is done in time step $T_1$ while keeping $g\simeq 0$. As shown in Fig.~\ref{fig_sup_singlet}, after this initial step all the relevant qubit states are below the first excited photon state. This configuration avoids undesired level crossings with higher photon number states during the next step of the protocol and only the $n=0$ manifold must be considered.  
  
During the second step $\omega_q^i\leq \omega_r/2$, but we still keep a finite frequency difference between the qubits to separate the state $\ket{\uparrow \uparrow \downarrow \downarrow}$ from other states with two qubits excited. This difference between the degenerate and non-degenerate qubit frequencies is visualized by  Fig.~\ref{fig_sup_singlet}(a) and (b). As the coupling $g$ is slowly increased while the difference in the qubit frequencies is tuned to zero, the state $\ket{\uparrow \uparrow \downarrow \downarrow}$ is adiabatically transformed into the $s=0$ state $|S\rangle$. During this process the state $|S\rangle$ become almost completely degenerate with the other $s=0$ state $|S^\prime\rangle$ [see Fig.~\ref{fig_sup_singlet}(b)]. However, also the non-adiabatic coupling between these two states is almost negligible, such that the preparation process is still adiabatic on the timescale of the protocol. 
    
In the last step of the protocol, the qubit frequencies are ramped up to the initial values as shown in Fig.~4 of the main text. At this point maintaining a frequency offset is not  crucial anymore. Note that in this last protocol step there are not restrictions on the operational time $T_3$ because the system is now in the dark state $|S\rangle$ and completely decoupled from the resonator mode.

\section{Disorder}
\begin{figure}
\includegraphics[width=0.65\textwidth]{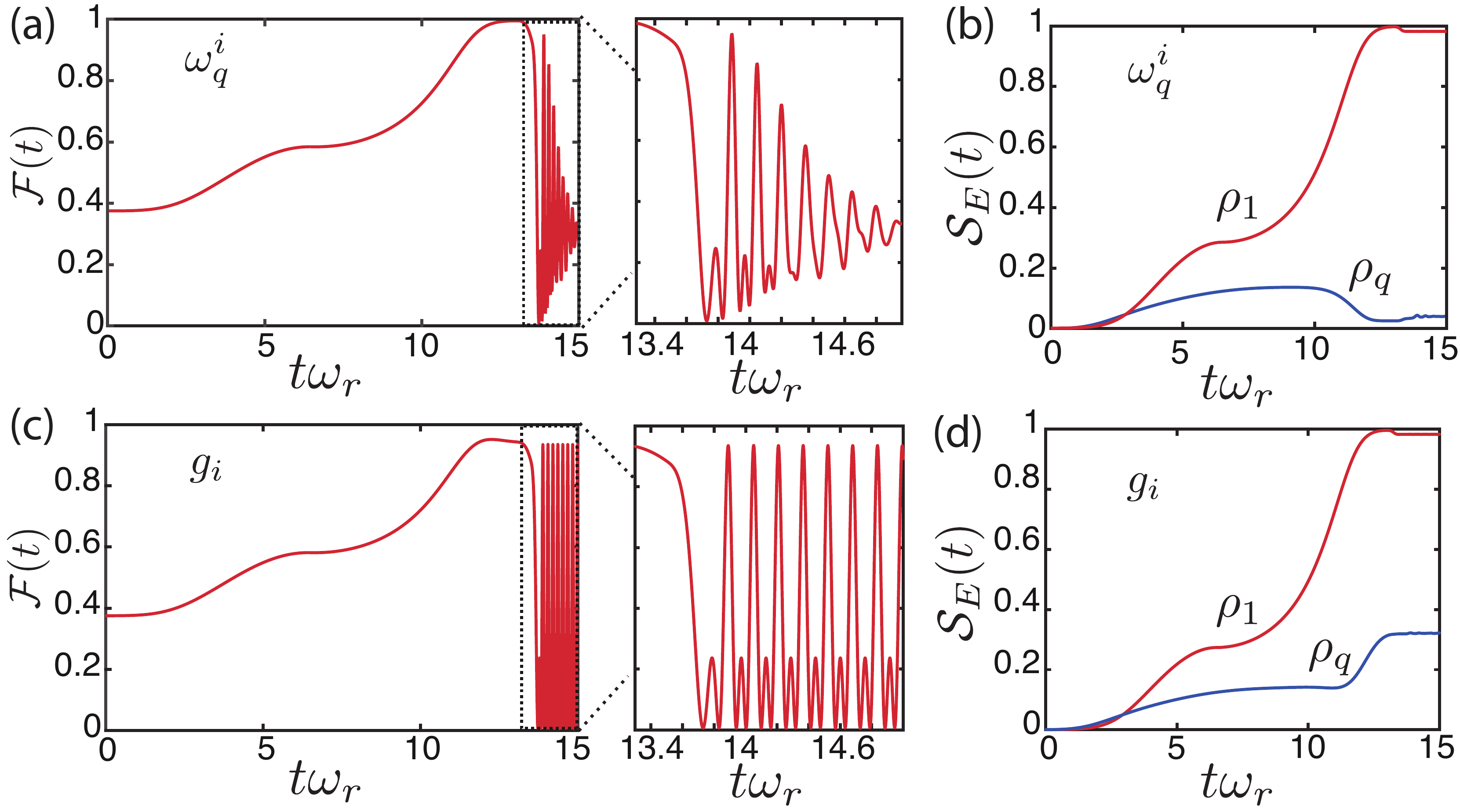}
\caption{(a)-(d) Fidelity and entanglement entropy as function of time in the presence of disorder obtained by averaging over $10$ simulation runs for $N=4$. In particular we show the entanglement entropy $S_E(t)=-\Tr\{\rho(t) \log_2(\rho(t))\}$ for the reduced density matrix of the qubit subsystem ($\rho_q(t)$) (blue line) and of a single qubit ($\rho_1(t)=\Tr_{N-1}\{\rho_q(t)\}$) (red line). In (a)-(b) the qubit frequency disorder is $\omega_q^i(t)=\omega_q(t)(1+\epsilon_i)$, while in (c)-(d) we have considered the coupling strength disorder $g^i(t)=g(t)(1+\epsilon_i)$, where $\epsilon_i$ are chosen randomly from a uniform distribution $[-0.1,0.1]$. All the other parameters for the protocol are as in Fig. 2(b) of the main text.}
\label{fig_sup_dis}
\end{figure}

In the main text we have assumed that our qubits are perfectly identical with matching frequencies, and that each of them couples to the resonator with the same coupling constant. However, due to fabrication disorder and control imprecisions this assumption can be difficult to achieve in experiments with multiple qubits. To evaluate the influence of disorder on the entanglement-harvesting protocol, we show in Fig.~\ref{fig_sup_dis} the results obtained from numerical simulations of the protocol in the presence of frequency and coupling disorder. Figure~\ref{fig_sup_dis}(a) shows the average fidelity, assuming that in each run of the protocol the individual qubit frequencies evolve as $\omega_q^i(t)= \omega_q(t)(1+\epsilon_i)$, where $\omega_q(t)$ follows the ideal pulse given in Fig. 2 of the main text and the $\epsilon_i$ are randomly chosen from the interval $[-0.1,0.1]$. We see that the main part of the protocol is essentially unaffected by frequency disorder, since the system is initially in the ground state and in the USC regime the system is dominated by the interaction terms. Frequency disorder only becomes important in the final decoupled state, where it dephases the symmetric state $\ket{D_0}$. Note, however, that for a fixed frequency distribution, this dephasing can be undone, since as shown in Fig.~\ref{fig_sup_dis}(a), it leads to almost no degradation of the purity or the degree of entanglement of the qubit state. In Fig.~\ref{fig_sup_dis}(c) and (d) the same plots are shown for the case of coupling disorder $g_i(t)=g(t)(1+\epsilon_i)$. Although, this type of disorder has a stronger influence on the evolution of the state, the plot shows that our protocol does not require strictly identical couplings and variation of around 10\% still lead to EEF $\mathcal{F}_{\rm E}\gtrsim 0.9$ and almost no degradation of the qubit-qubit entanglement. The main quantity affected is the entanglement entropy of the qubit subsystem, which does not approach the value of zero, thus showing that qubits and resonator are not perfectly decoupled. However, we note that this measure of entanglement is very sensitive in our case, since the qubit state we achieve at the end of the protocol coincides with our target state with fidelity above $90\%$.

\section{Implementation of the protocol}
\begin{figure}[!htbp]
    \begin{center}
        \includegraphics[width=0.25\textwidth]{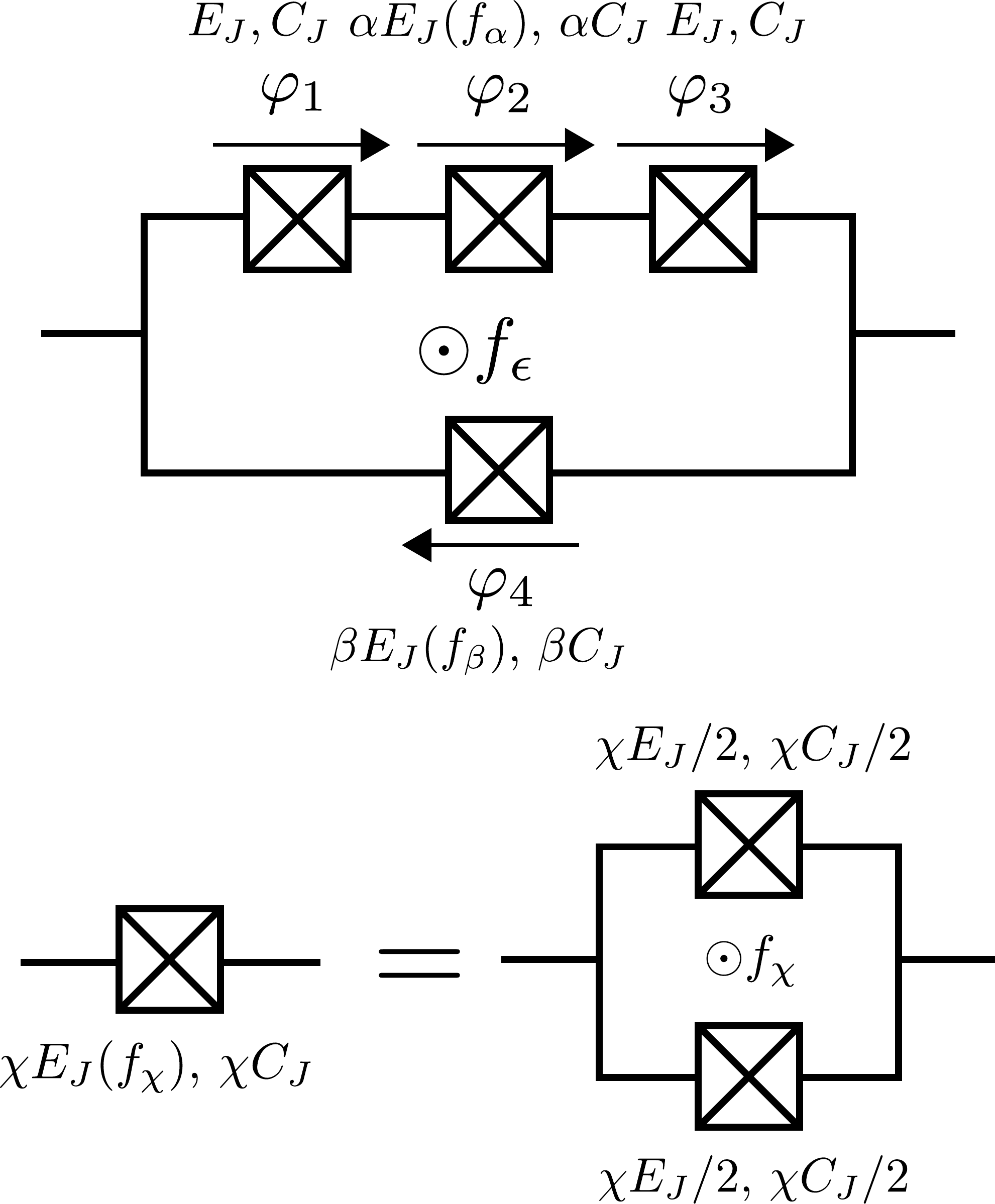}
        \caption{Flux-qubit circuits considered for the implementation of  tunable qubit frequencies and qubit-resonator couplings. The SQUID-loops behave as an effective junction with a flux-tunable Josephson energy.}
        \label{Fig6}
    \end{center}
\end{figure}

In this section we describe a specific flux-qubit circuit, which can be operated in the USC regime and allows a high tunability of the qubit frequencies and couplings. We propose to achieve this goal by using four-junction flux qubits~\cite{Qiu2016s} with two of the junctions replaced by a SQUID-loop, effectively turning them into junctions with a flux-tunable Josephson energy. The flux qubit design is depicted in Fig.~\ref{Fig6}.
The tunable junctions are junctions 2 and 4 which have sizes $ \alpha $ and $ \beta $, respectively, with respect to junctions 1 and 3. We denote by $ \varphi_i = \Phi_i/\Phi_0 $ the jump of the superconducting phase across the junction $ i $, where $ \Phi_0 = \hbar/2e $ is the reduced flux quantum and $ \Phi_i $ the jump of the generalized flux across the junction. The conjugate charge to $ \varphi_i $ is denoted as $ n_i $. The phases $\varphi_i$ are not independent but constrained by the flux quantization condition for the three loops (the big loop and the two SQUID-loops $ \alpha $ and $ \beta $)
\begin{align}
    \sum_{i \in \{1,2,3,4\}} \varphi_i + f_{\epsilon} &= 0,\nonumber\\
    \sum_{i \in \{2,6\}} \varphi_i + f_{\alpha} &= 0,\\
    \sum_{i \in \{4,5\}} \varphi_i + f_{\beta} &= 0\nonumber,
\end{align}
where $ f_{\eta} = \Phi_{\eta}/\Phi_0 $ and $ \eta = \alpha,\,\beta,\,\epsilon $ is the magnetic frustration through the loop created by external magnetic fluxes $ \Phi_{\eta} $. Using the above equations we eliminate phase jumps $ \varphi_{2} $, $ \varphi_{5} $ and $ \varphi_{6} $ from the problem. The standard quantization procedure for circuits then gives the Hamiltonian \cite{vool16s,Qiu2016s}
\begin{align}
    H_q &= \dfrac{4E_C}{\alpha + \beta + 2\alpha\beta} \left[ (\alpha + \beta + \alpha\beta) (n_1^2 + n_3^2) + (1 + 2\alpha)n_4^2 - 2\alpha\beta n_1n_3 - 2\alpha(n_1 + n_3)n_4 \right]\\
    &\qquad- E_J \left[ \cos(\varphi_1) + \alpha\cos\left(\dfrac{f_{\alpha}}{2}\right)\cos(\varphi_1 + \varphi_3 + \tilde{\varphi}_4 + \tilde{f}_{\epsilon}) + \cos(\varphi_3) + \beta\cos\left(\dfrac{f_{\beta}}{2}\right)\cos(\tilde{\varphi}_4) \right]\nonumber,
\end{align}
where $ \tilde{\varphi}_4 = \varphi_4 - f_{\beta}/2 $ and $ \tilde{f}_{\epsilon} = f_{\epsilon} + (f_{\beta} - f_{\alpha})/2 $. From the shape of the Hamiltonian we can see that if we tune the frustration parameters, $ f_{\alpha} $, $ f_{\beta} $ and $ f_{\epsilon} $, in unison such that $ f_{\epsilon} = (2\pi + f_{\alpha} - f_{\beta})/2 $, the structure of the Hamiltonian stays the same except that the effective Josephson energy of the SQUID-loops vary sinusoidally with the frustration. This enables us to operate the flux qubits at the sweet spot, $ \tilde{f}_{\epsilon} = \pi $, while changing the potential landscape. Note that in practice the cross-talk between the magnetic fluxes may complicate the qubit control, but in principle it is always possible to measure this cross-talk and compensate it by appropriately chosen control pulses.

The flux qubit couples to the resonator through the phase jump over the entire qubit (see the main text), which, with our notation, is given by $ \Delta\varphi = \tilde{\varphi}_4 $. The coupling constant $ g $ between the resonator and the qubits is proportional to the matrix element of $ \Delta\varphi $ between the ground and excited states of the qubits, $ \Delta\varphi_{eg} = \langle e | \Delta\varphi | g \rangle $. Additionally, the coupling to the resonator renormalizes the qubit Hamiltonian by adding a term $ E_L\Delta\varphi^2/2 $, where $ E_L = \Phi_0^2/L $ is the inductive energy related to the resonator inductance $ L $, to the qubit Hamiltonian.

Now we are ready to demonstrate the tunability of the qubit frequency and qubit-resonator coupling. We diagonalize the qubit Hamiltonian $ H_q $, plus the renormalization term coming from the coupling, numerically to find the eigenfrequencies and evaluate the transition matrix element $ \Delta\varphi_{eg} $. We choose the following parameters for the simulation: $ \alpha = 0.6 $, $ \beta = 6 $, $ E_L/h = 2.57\,\mathrm{GHz} $, $ E_C/h = 4.99\,\mathrm{GHz} $ and $ E_J/h = 99.7\,\mathrm{GHz} $. Our choice of $ E_L $ sets the resonator inductance to $ L = 63.7\,\mathrm{nH} $. In addition we choose $ C = 1.59\,\mathrm{pF} $ which determines the resonator frequency and impedance to be $ \omega_r = (LC)^{-1/2} =  2\pi\times500\,\mathrm{MHz} $ and $ Z_r = \sqrt{L/C} = 200\,\Omega $, respectively. In the simulation we tune $ f_{\alpha} $ from $ 0 $ to $ 0.70\pi $ and $ f_{\beta} $ from $ 0 $ to $ 0.96\pi $ ($ f_{\epsilon} $ changes accordingly to keep the qubit at its sweet spot).

\begin{figure}[!htbp]
    \centering
    \includegraphics[width=0.8\textwidth]{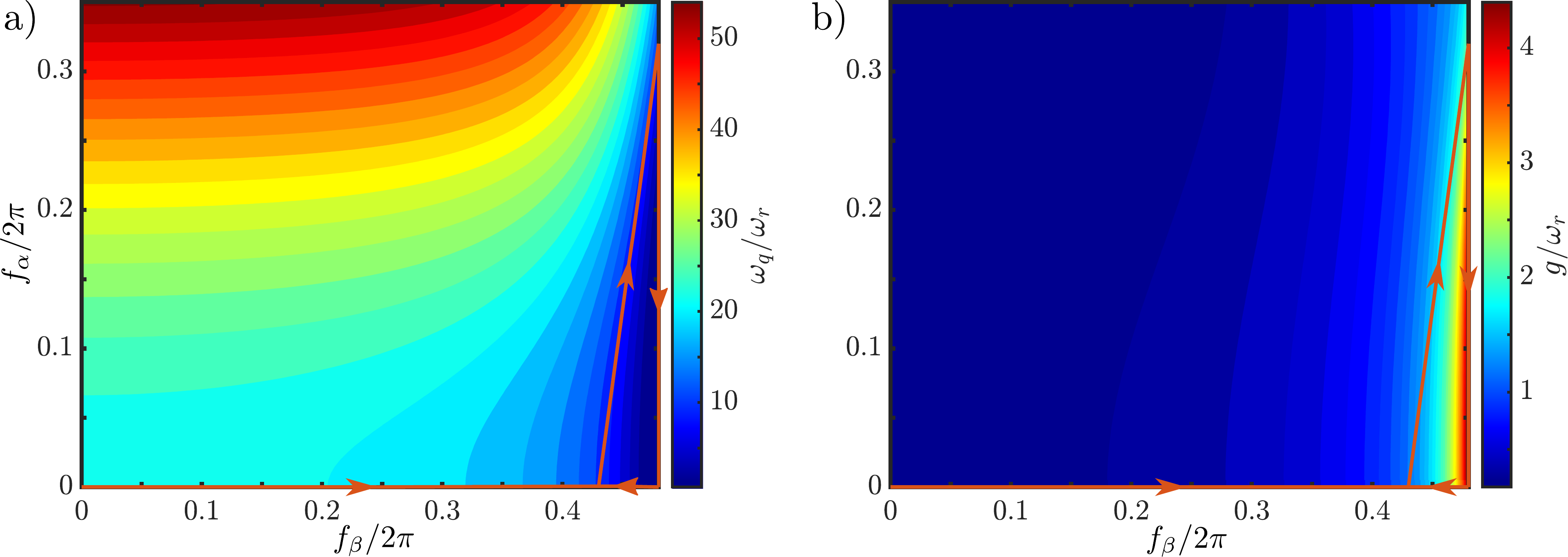}
    \caption{Tunability of the qubit. a) Transition frequency of the qubit, $ \omega_q $, as a function of the external fluxes in the SQUID-loops in units of the resonator frequency. b) Normalized coupling constant $ g/\omega_r $ of the qubit to the $ LC $-resonator. Parameters used to produce this plot are given in the text.}
    \label{Fig7}
\end{figure}

\begin{figure}[!htbp]
    \centering
    \includegraphics[width=0.8\textwidth]{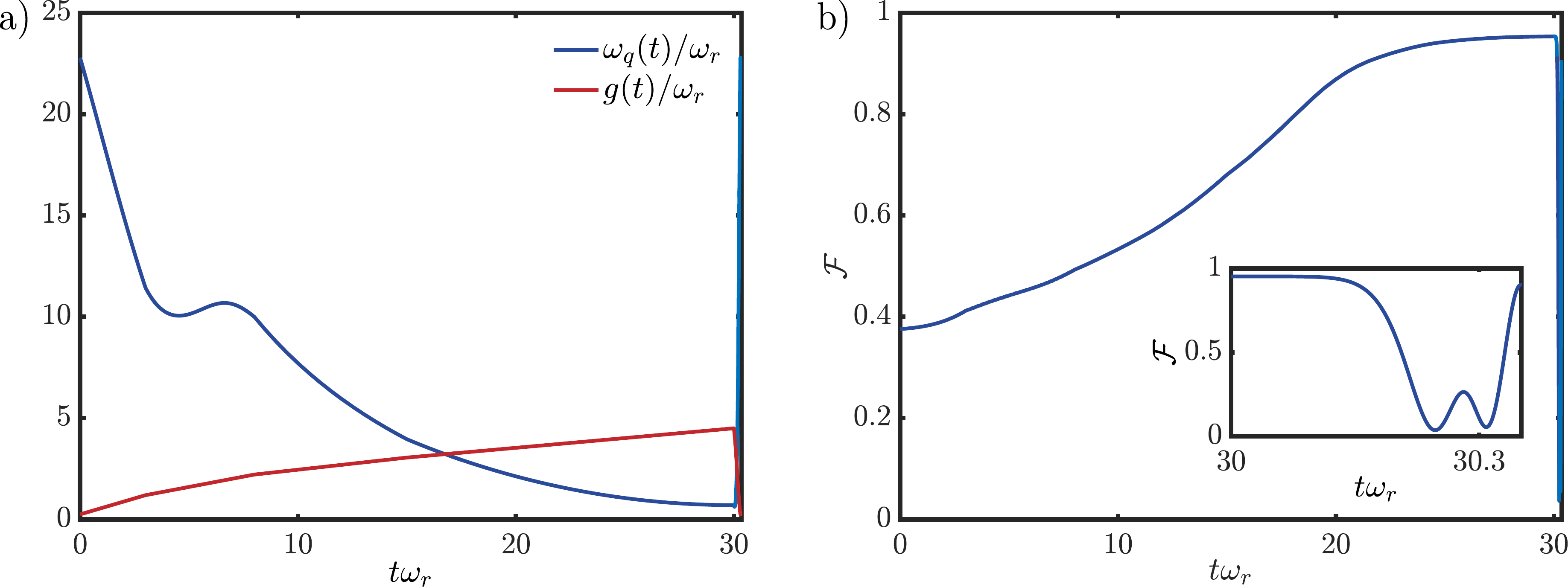}
    \caption{a) Pulse shapes for the parameters $ g $ and $ \omega_q $ obtained from following the path  outlined in Fig.~\ref{Fig7} for the external control parameters. b) Fidelity of the protocol for $N=4$. The inset shows the final stage of the protocol.}
    \label{Fig8}
\end{figure}

In Fig. \ref{Fig7} a) we plot the transition frequency of the qubit, normalized to the resonator frequency against the external fluxes in the two SQUID-loops. The qubit frequency is highly tunable ranging from $ \sim 50 \omega_r $ all the way to $ \sim 0.5 \omega_r $. The tunability of the normalized coupling constant, $ g/\omega_r $, is demonstrated in Fig. \ref{Fig7} b). It is also highly tunable ranging from $ \sim 0.17 $ up to $ \sim 4.5 $. The only limitation we have here is that the qubit transition frequency and qubit resonator coupling cannot be tuned entirely separately. The path that we propose to take in the $ (f_{\alpha},\, f_{\beta}) $ landscape is portrayed in Fig. \ref{Fig7} with the red curves. We start from the point $ (0,\, 0) $ and then traverse the curve clockwise, as the arrows show, back to the origin. In Fig. \ref{Fig8} a) we plot the proposed time-dependent pulses for $ g(t) $ and $ \omega_q(t) $. We have chosen the shape of the pulse for $ g $ and from that computed the pulse shape of $ \omega_q $. The pulse sequence consists of two parts: ramping up $ g/\omega_r $ from $ 0.25 $ to $ 4.5 $ adiabatically and then tuning it back down to its initial value non-adiabatically. The qubit frequency starts from $ 22.8\omega_r $, goes down to $ 0.7\omega_r $ and then ramps back up to its initial value. Fig. \ref{Fig8} b) shows the fidelity obtained with the pulse of a) in the case of four qubits coupled to the LC-resonator. Even with a limited individual tunability of $ g $ and $ \omega_q $ we obtain fidelities of $ \mathcal{F}_{\rm E} > 0.9 $. For two qubits we can obtain a fidelity of $ \mathcal{F}_{\rm E} \approx 0.96 $ with the same pulse.

\section{Influence of higher resonator modes}
In our analysis in the main text we have assumed a single mode lumped-element resonator and have neglected the influence of all higher modes. For a typical lumped-element resonator of dimension $d=500\mu$m and a resonance frequency of $2\pi\times 5$ GHz, the next higher mode is estimated to be at $\omega_e\approx c/(6d)\sim2\pi\times 100$ GHz. In Ref.~\cite{Jaako2016s} we have shown that for such a high ratio $\omega_e/\omega_r > 20$, the effect of a higher circuit mode has no relevant influence on the resulting USC physics. By simply scaling this circuit by a factor $10$, we would obtain a fundamental resonance of $\omega_r/2\pi\approx 500$ MHz and an excited mode at around $\omega_e/2\pi\approx 10$ GHz. By using an optimized design to increase this frequency or simply changing the maximal qubit frequency to, e.g., $8$ GHz would avoid a resonant coupling to this mode without affecting the protocol. 

During the first stage of the protocol the system is in the ground state. There might be some weak admixing with the excited mode $\sim(g/\omega_e)^2$, which however, should not affect the adiabaticity condition. During the USC part of the protocol, the qubit frequency is tuned to a very low value. In this case the relation between $g$, $\omega_q$ and $\omega_r$ is very similar to the values assumed in the analysis in Ref.~\cite{Jaako2016s}. Therefore, according to this study the system properties should not be considerably affected. Finally, during the last stage of the protocol, where the qubit frequency is ramped-up again, the coupling has already been switched back to a very small value. The mixing with the excited mode will be small, $\sim g^2_{\rm min}/(\omega_e-\omega_q)^2$. 

In summary, based on these estimates we do not expect a significant degrading of our protocol from interactions with higher-order modes of the lumped-element resonator.

\section{Numerical simulations}
\subsection{Coherent evolution}
In this short paragraph we provide some details about the numerical simulations, which have been used to produce the plots in the main text.

For the plot of the eigenvalues in Fig. 1 in the main text we have diagonalized Hamiltonian (3) using a truncated set of $140$ number states for the resonator mode. For the implementation of the entanglement harvesting protocols shown in Figs. 2, 3 and 4 in the main text we have numerically integrated the time-dependent Schr\"odinger equation using $100$ resonator states. In all calculations we have verified that increasing this number of basis states does not affect our results. 

In Fig. 3(c) in the main text we plot the EEF for $N = 4$ and for a resonator mode initially in a thermal state $\rho_{\rm th}=\sum_n p(n)\ket{n}\bra{n}$ with $p(n)$ the Gibbs distribution at temperature $T$. The fidelity has been obtained by solving the Schr\"odinger equation for each initial number state separately, and then averaging all the resulting fidelities according to the thermal probabilities, $p_n$.
For this plot we have included the first $10$ resonator Fock-states and verified that averaging over a thermal distribution including more resonator states does not change the result.

\subsection{Master equation simulations}
To make sure dissipation during the protocol does not spoil our scheme we perform simulations including cavity decay into the dynamics. We do so by modelling the coupling to the bath by a Markovian master equation. Note, however,  that in the ultrastrong coupling regime the dissipator is not given by the photon annihilation operator $ a $ and must be expressed in terms of the coupled eigenbasis states $|k\rangle$  of the system Hamiltonian~\cite{beaudoin11s, ridolfo12s}. For the case of the extended Dicke model Hamiltonian and assuming an ohmic spectral density of the bath, the resulting master equation reads
\begin{figure}[h!]
    \centering
    \includegraphics[width=0.8\textwidth]{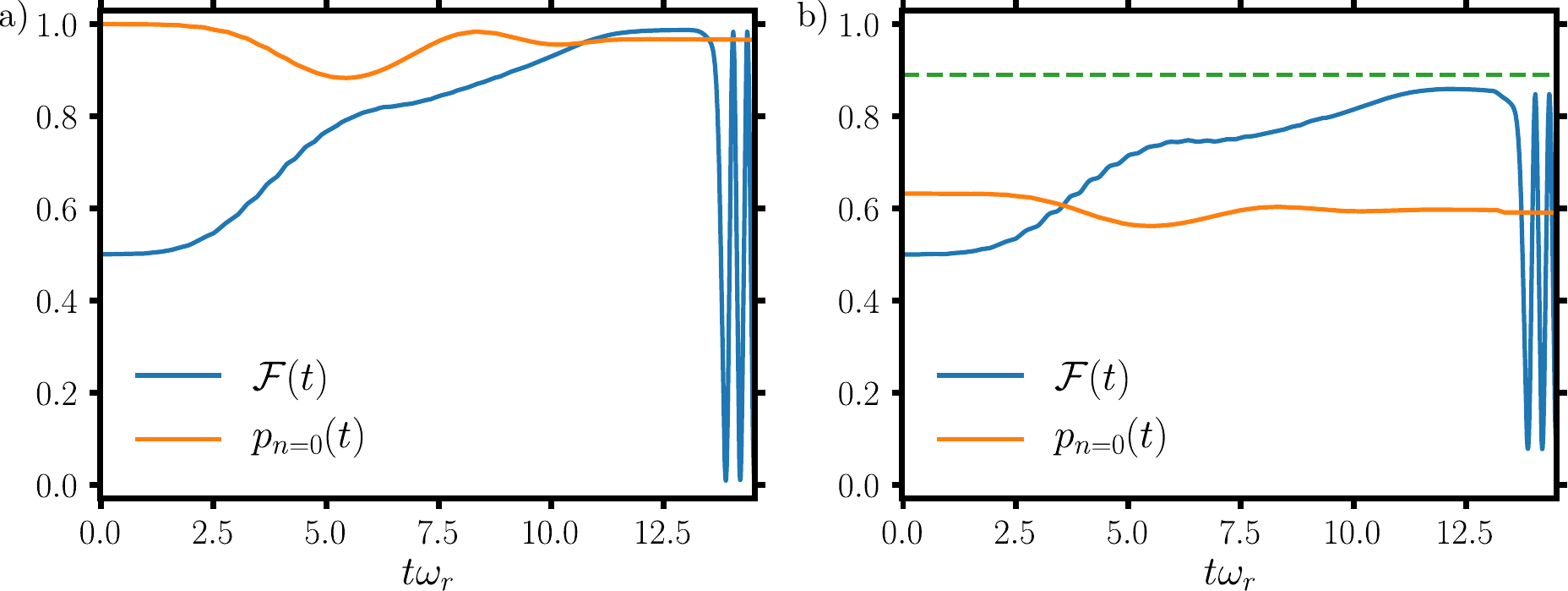}
    \caption{Fidelity $\mathcal{F}(t)$ of the protocol for two qubits in the presence of cavity decay for $ Q = 100 $ and $ \omega_{\rm max} = 10\omega_r $. Other parameters used are the same as in Fig. 2 of the main text. a) $ T = 0 $. b) $ T = \hbar \omega_r/k_B $. Dashed green line indicates the achievable fidelity without cavity decay at the same temperature.}
    \label{Fig9}
\end{figure}
\begin{align}\label{ME}
    \dot{\rho} = -\dfrac{i}{\hbar}[H, \rho] + \sum_{k,\, l > k} \Gamma_{kl}(1 + \bar{n}(\Delta_{lk}, T)) \mathcal{D}[\ket{k}\bra{l}]\rho + \sum_{k,\, l > k} \Gamma_{kl} \bar{n}(\Delta_{lk}, T)) \mathcal{D}[\ket{l}\bra{k}]\rho,
\end{align}
where $ \Gamma_{kl} = \kappa\dfrac{\Delta_{lk}}{\omega_r} |C^{a}_{kl}|^2 $, $ \Delta_{lk}=E_l-E_k$  is the energy difference between eigenstates $l$ and $k$, $ C^{a}_{kl} = \bra{k} a + a^{\dagger} \ket{l} $ and $ \kappa = \omega_r/Q $ is the resonator decay rate in the weak coupling regime. $ \bar{n}(\Delta_{lk}, T) $ is the occupation of the environmental modes at the transition frequency $ \Delta_{lk} $ at temperature $ T $. The superoperator $\mathcal{D}$ is defined in the standard way as $ \mathcal{D}[O]\bullet =(2O \bullet O^{\dagger} - \{ O^{\dagger}O,\, \bullet \})/2 $. Because we change $ \omega_q $ and $ g $ in time the basis $ \{ \ket{k} \} $, and thus also the jump operators and rates, changes in time. 
For the numerical simulation of Eq.~\eqref{ME} we diagonalize the system Hamiltonian at each point in time to construct the correct jump operators and transition rates during the different stages of the protocol. After constructing the correct operators and transition rates we move back to the composite basis $ \ket{n, s, m_z} $ spanned by the Fock states $ \ket{n} $ and collective spin states $ \ket{s, m_z} $. This is done with a time-dependent unitary transformation $ U(t) = \sum_{k,\,n,\,m_z} \ket{n, s, m_z}\bra{k(t)} $, where $ \ket{k(t)} $ is the instantaneous eigenstate of the extended Dicke Hamiltonian.


Using the above prescription we simulate our protocol in presence of photon decay for a two qubit system. In Fig. \ref{Fig9} we plot the fidelity for an LC resonator of $ Q $-factor $Q=100 $ in contact with a bath at temperature $ T = 0 $, \ref{Fig9}a), and $ T = \hbar\omega_r/k_B $, \ref{Fig9}b). We see that even for such a low $ Q $-factor the cavity decay has no significant effect. At $ T = 0 $, where the system is essentially at all times in the ground state, the performance of the protocol is not affected at all and even at $ T = \hbar\omega_r/k_B $ the degradation is only on the level of $ \sim 3.5\% $. In Fig. \ref{Fig9}b) the dashed green line indicates the performance of the protocol at $ T = \hbar\omega_r/k_B $ without including cavity decay.

\end{document}